\documentclass[a4paper]{VisionStyle}
\usepackage{epsfig}

\begin{document}

\title{XMM-Newton Calibration - an  overview}

\author{D.H.\,Lumb\inst{1} } 

\institute{ESA Payload Technology Division, Research and Scientific Support Department, ESTEC, Postbus 299, NL-2200 AG Noordwijk, The Netherlands}

\maketitle 

\begin{abstract}
This paper introduces the articles that describe detailed aspects of the XMM-Newton 
calibration. The unique calibration issues of XMM-Newton are highlighted. The original 
calibration requirements and aspects of the ground calibration are summarized. 
The life cycle of the in-orbit calibration observations, analysis and ingestion into 
calibration files is discussed

\end{abstract}

\section{The original calibration requirements}\label{dlumb-WA2_sec:reqs}

The mission science goals were defined more than 15 years before launch. During the
hardware development phase these were used to form a set of requirements for calibration
accuracy that guided the design of facilities and activities for the instrument
calibration (\cite{dlumb-WA2:scd}). With the benefit of hindsight it was realised that the change in some aspects
of projected 
performance of the delivered hardware undermined or superceded some of those goals. For example the
significantly improved telescope resolution demands much greater astrometric reconstruction
and PSF knowledge than was envisaged some years ago, when it was by no means certain the
goal of 20 -- 30 arcsecs Half Energy Width would be met! Nevertheless those sets of requirements
have remained broadly appropriate (see Table~\ref{dlumb-WA2_tab:tab1}), and it is instructive to recall them in order to judge
the success of our existing activities.

\begin{table*}[bht]
  \caption{Summary of major calibration requirements}
  \label{dlumb-WA2_tab:tab1}
  \begin{center}
    \leavevmode
    \footnotesize
    \begin{tabular}[h]{lcll}
      \hline \\[-5pt]
      Parameter & Requirement     &  Notes&Achieved\\[+5pt]
      \hline \\[-5pt]     
Astrometry &$\leq$3 arcsec&Allows ground based follow-up&1 arcsec if field\\
&&with low confusion probability&source IDs available\\
&&&\\
Absolute eff. area &10\%&Long term variability studies&5 -- 10\% (0.5 -- 7~keV)\\
&&Observatory cross-calibrations&\\
&&&\\
Relative eff. area &3\%&Robust spectral fitting&Typically2\%, 5\% at edges? \\
&&&\\
Energy calibration& 3eV EPIC,   &Broadening and bulk motions&Not achieved for all modes\\
& better than 10mA RGS&limited by statistics only&Some bright SNR require $\leq$3eV\\ 
&&&\\
Timing&1 msec&Fast CCD readouts would allow&Internally 10$\mu$sec\\
&&spectrally resolved light curves&Limited by Orbit knowledge TBC\\
&&&\\
OM photometry &10  mMag&Populations and variability&Relative stability OK\\
&&&Completing the colour transforms per filter\\
&&&\\
OM astrometry & 1 arcsec&Source ID in crowded fields&1 arcsec if field\\
&&&source IDs available\\      \hline \\
      \end{tabular}
  \end{center}
\end{table*}

At the same time there was a widely held opinion that there was no need to calibrate 
XMM on-ground to such accuracies, because XMM would be contemporaneous with an AXAF 
observatory which would be ''{\em calibrated to 1\%}'', and therefore XMM needed only to be
cross-calibrated to AXAF. Needless to say that approach was not adopted.

\section{Status at Launch \& CAL/PV phase}
\label{dlumb-WA2_sec:launch}
The usual realities of the flight hardware programme contributed to a lowering
of expectations, not least for the calibration accuracy. As ever there was not
enough time to complete the designed calibration measurement programme. 

One problem was that of
late deliveries. At a late stage the PN camera Flight Model suffered a failed CCD, and
was replaced with the Flight Spare camera (which eventually turned out to be physically 
different from the original). There was a need to replace some of the CCDs in the MOS
cameras, even swapping one complete focal plane. The OM began a late replacement 
programme for its mirrors, leading to a limited test programme time, which
affected the calibration of filters and stray light testing. Overall the
in-flight calibration campaign did not have the sound under-pinning of a successful
ground calibration programme that had been anticipated. On the other hand, the
multiple similar instrument deployment approach {\em has} meant that there
is strong heritage from one version of a camera to the other which recovers
some of this loss.  

The in-flight calibration programme was scoped in quite some detail somewhat before
launch in order to ensure that an efficient completion of the complex measurement programme
could be planned for. However, it  was not known if the Ground Segment would support this 
plan, as it was never the subject of end-to-end tests before launch. As a result the
celestial calibration programme was severly curtailed by
the eventual low efficiency of operations in the early mission phase. Such aspects
as needing to develop functionality (Observation Data File availability, missing
Atitude History Files etc.), as well as developing workarounds and changing 
the instrument
operations and settings to accommodate instrument anomalies, were all
carried out during  the
nominal calibration phase. 

In reality then, we have been faced with building up the calibration knowledge
during the course of normal observations of GT and GO targets, so that the availability
of improving knowledge lags behind the expectations.

The internal instrument in-flight calibration sources have limited energy leverage, so that their 
use for confirming energy scales and understanding effects of gain stability is less
than anticipated. Most efforts therefore rely on celestial targets, none of
which could be described as ``{\em standard candles}''.

Despite all these difficulties, the instrument teams have laboured to develop an understanding
of the instrument calibration, whose knowledge at one year post-launch is captured in the
accompanying documents, and which it should be judged, compares favourably with the original
goals laid out. 

\section{Uniques features of XMM-Newton}
\label{dlumb-WA2_sec:unique}
It is perhaps worth reiterating some of the exceptional features of the XMM-Newton payload
which pose some very particular challenges for the calibration activity. \begin{itemize}
\item The unprecedented large effective area means that the photon statistics imply 
systematic errors are revealed BEFORE they would be in other observatories. Also the improved S:N pushes new 
astrophysics models, necessitating better calibration
\item The co-aligned instruments operate simultaneously - while this in principle allows 
to cross-check between instruments, it also demands from observers that the 
cross-calibration is secure
\item Filter choices: EPIC for example can select from THIN, MEDIUM or THICK,  and
OM has multiple positions for filter photometry. These require a multiplicity in calibration
activity and understanding
\item Instrument modes: there are different modes for count rate optimisation : EPIC 
offers IMAGING,WINDOW and TIMING modes. The OM also offers fast and  windowed operation
to its nominal mode. Again a multiplicity of calibration is imposed, compounded
by the need to use windowed modes in these instruments for bright high S:N sources.
\end{itemize}

\section{Calibration life-cycle}
\label{dlumb-WA2_sec:cycle}
There are scheduled routine and periodic calibrations to check the stability of gain, 
wavelength scale, CTI, astrometry etc.. Thereafter we define non-routine observations to 
investigate anomalies. The mission planning cycle is a first limit to the speed of 
implementation, especially where visibility of a chosen target may be limited. Recent experience shows
anODF will become available $\leq$1 month after the observation. The instrument teams 
then analyse this data. If the improved understanding or models are forthcoming, it can frequently
occur that new interfaces in  SAS tasks or via. the ''{\em Calibration Access Layer}''
may be needed. Such new tasks require that adequate testing will occur before the implementation in the next public
release of SAS. Thus it is easy to comprehend the low efficiency of the cycle of updates that sometimes occurs.

As well as the activity in the instrument teams, the calibration effort is supported by the SOC team who man the Help Desk and enter the CCF files and 
generate the Release Notes explaining the changes to the calibration data.
\subsection{Current Calibration File}
\label{dlumb-WA2_sec:ccf}
In the simplest cases, the review of calibration analysis leads only to an improved knowledge of
exiting calibration models, and only a data set update is necessary. In the XM-Newton
SAS environment these calibration sets are known as the "{\em Current Calibration File}".

Release notes are published with each new CCF set to explain what has changed and 
the science impact. This means that the user can judge if his/her data needs
to be reprocessed.
The $CALVIEW$ task can be used to plot out and export data as interpreted by SAS 
Calibration Access Layer calls. This is an additional useful aid for understanding
the impact of various calibration data.

Users should be aware of the $CIFBUILD$ task, and ensure to run this task to get 
complete new CCF set aligned for reprocessing of data sets. It is also possible to
run individual tasks with the  $-ccffiles "xxxx.CCF"$ option, just to
see the effect on one task, of a changed calibration file.'$xxxx.CCF$' represents
the full name and path of a single calibration file
Finally we recommend that users
subscribe to the calibration mailing list for prompt notification of CCF updates.  

\section{Outlook}
The existing calibration is clearly adequate to support a host of science
papers that are now being produced from XMM observations. Cross-calibration
between the instruments of XMM-Newton  is excellent with the exception of the extreme
energy ranges. As noted by Snowden in this volume, there is also an acceptable
cross calibration between EPIC and other observatories. The status
of the knowledge presented herein is expected to be accounted for in the next 
public release of SAS (version 5.3 at time of writing). 

However it was evident from some presentations at this workshop that new science analysis 
(for example subtle effects in the iron line details in AGN) afforded by
the unrivalled XMM-Newton collection area are demanding yet higher calibration
accuracy. The instrument teams have committed significant teams of expertise
to support the continued improvement in understanding the calibration of their
hardware, so that we expect a continual evolution in this knowledge through the
extended mission duration.

\end{document}